\documentclass[conference]{IEEEtran}
\usepackage{inputenc}
\usepackage[english]{babel}
\usepackage{amsmath}
\usepackage{amsfonts}
\usepackage{amssymb}
\usepackage{graphicx}
\usepackage{hyperref}
\usepackage{listings}
\usepackage{rotating}

\begin{document}

\title{Towards digitalisation of summative and formative assessments in academic teaching of statistics}

\author{Nils Schwinning, Michael Striewe, Till Massing, Christoph Hanck, Michael Goedicke \\
University of Duisburg-Essen\\
Essen, Germany\\
\{nils.schwinning,michael.striewe,till.massing,christoph.hanck,michael.goedicke\}@uni-due.de}

\maketitle

\begin{abstract}
Web-based systems for assessment or homework are commonly used in many different domains.
Several studies show that these systems can have positive effects on learning outcomes.
Many research efforts also have made these systems quite flexible with respect to different item formats and exercise styles.
However, there is still a lack of support for complex exercises in several domains at university level.
Although there are systems that allow for quite sophisticated operations for generating exercise contents, 
there is less support for using similar operations for evaluating students' input and for feedback generation.
This paper elaborates on filling this gap in the specific case of statistics.
We present both the conceptional requirements for this specific domain as well as a fully implemented solution.
Furthermore, we report on using this solution for formative and summative assessments in lectures with large 
numbers of participants at a big university.
\end{abstract}

\section{Introduction}
Using computer assisted assessment (CAA) to face growing numbers of participants in university courses is a well-known concept.
Especially web-based homework systems are very popular given decreasing numbers of teaching assistants that are able to grade 
homework manually. Studies show that using web-based homework instead of paper-based concepts does not lead to decreases in students 
performances.\cite{Lenz2010WebBasedHomework}\cite{HerronEtAl2012SuccessRatesComparison}\cite{GokEtAl2011ComparisonOfStudentPerformancePhysics}.
On the contrary, a majority of studies is able to report a positive effect of using computer-assisted instruction in the classroom specifically for the domain of statistics \cite{SosaEtAl2001EffectivenessofCAI}.

Besides being useful for instruction and homework, some systems are even able to perform summative assessments such as tests or exams which reduces the effort 
of required manual grading even further. Many systems exist that offer the classical digital exercise formats such as 
multiple choice, fill-in or drop-down. Examples include Moodle\footnote{\url{https://moodle.org/}}, 
LonCapa\footnote{\url{http://lon-capa.org/}} and Stack\footnote{\url{https://stack.maths.ed.ac.uk/demo/}} among others.
Sometimes, these formats can be combined to create an exercise consisting of several subtasks or the systems allow to create 
exercises with variable content by using randomly created elements.

However, courses in higher statistics need to use more sophisticated exercise designs that cannot be easily transferred to a such system.
Many of the typical assessments in the area consist of open tasks where students have to choose a certain strategy in order to solve the task correctly.
A very important question is, how these exercises can be digitalized without losing quality, which means to ask questions
without implicitly revealing the solution strategy. Therefore, the system to be used needs to allow a most flexible exercise design, 
that enables authors to react adequately to submissions. This applies both for giving detailed feedback messages as well as for using 
wrong solutions for further calculations. Furthermore, additional advantages of CAA can be used to help students to increase their learning 
outcome. In particular, it would be beneficial to use parameterized exercises which give students the chance to work on the same problem several times. 

To be able to do all this, many complex statistical functions are needed. In order to avoid the effort of implementing these in an existing 
system it seems reasonable to consider the use of a tool that offers this functionality. The programming language \textsc{R}\footnote{\url{https://www.r-project.org/}} 
is a free software widely used in the area of statistics as it provides many built-in functions. Besides, its architecture makes it highly extensible
which means that user can install packages written by others or even write packages with their own functions. All these factors make it very attractive to use \textsc{R} together with CAA to realise a concept that allows the digitalisation of exercises for higher statistics.

This paper presents an approach to combine the requirements worked out above into an exercise type that allows us to digitalize exercises 
for higher mathematics using the example of statistics and thus to support university courses with large numbers of participants. 
To do this, we have made some major improvements to the e-assessment system JACK which is in use at the University of Duisburg-Essen. 
A central new feature is the interaction with the statistical programming language \textsc{R} in exercises. This means 
that \textsc{R} cannot only be used for creating random parameters at the beginning of an exercise but we can also send 
submissions to it and use it for evaluations. Furthermore, it is possible to evaluate submitted expressions with \textsc{R} 
and use these values in subsequent tasks of an exercise. We use the package \textsc{Rserve}\footnote{\url{https://rforge.net/Rserve/}} 
that sets up a TCP/IP server which can be easily used from various programming languages such as \textsc{Java} or \textsc{C/C++}
to connect our system to \textsc{R}. Through this setup the approach can be transferred to various mathematical subdomains using other external
systems with minimal effort. The concept is in heavy use in statistics lectures with large numbers of participants at our university and is used for both, 
formative and summative assessment types.   

The remainder of this paper is organized as follows:
Section \ref{sec:related_work} provides a brief review of existing approaches
for the same problem and reports on their individual drawbacks in our context.
Section \ref{sec:central_message} discusses the features that we equipped our exercise type 
with in order to use it in introductory courses at big universities, followed by some examples.
A short description of the technical realization is given in section \ref{sec:realization}. 
Experiences from practical use of the system are reported in section \ref{sec:in_use} before the paper is
closed with conclusions and future work in section \ref{sec:conclusions}.  


\section{Related Work}
\label{sec:related_work}
\subsection{Tool Support for Teaching Statistics}
One of the arguments for using electronic systems in the statistics classroom is
to create higher involvements of students and thus to stimulate learning. A
dedicated experiment on using a personal response system (PRS) as additional
tool in a statistics lecture is presented in
\cite{Wit2003AudienceResponseSystem} and reports both benefits and drawbacks.
One specific limitation mentioned is the fact that formal exams in the course
use open questions, while the PRS is based on multiple-choice questions.

A way to overcome this gap is to make use of another arguments in favour of
digitalisation of exercises: Tool support can be used to generate different
variants of the same (open-ended) exercise. The most direct way to do this for
exercises in higher statistics is to use \textsc{R} in combination with the
package \texttt{exams}. This package is able to generate exercises from
templates producing output in PDF, HTML, or specific formats for some common
e-learning-systems
\cite{GruenZeileis2009AutomaticGeneration,ZeileisEtAl2012FlexibleGeneration}.
While this allows authors to use the full power of \textsc{R}, it comes with two
major drawbacks with respect to our goals: First, the results are static, thus a
student attending a specific exercises in an e-learning-system will always get
the same contents and has hence no benefit from using \textsc{R} in the
generation process. Second, \textsc{R} is not involved in the evaluation of
student input and hence also not involved in the generation of feedback or
hints.

A different approach which is able to use the same power of calculations both
for exercise content generation and evaluation has been realized by the
\textsc{CAMPUS} project \cite{Hunt1998CAMPUS} almost 20 years ago. It is
entirely based on Microsoft Excel spreadsheets and incorporates the use of the
\texttt{RAND()} function to generate random instances of exercises. While this
is an elegant solution that is both useful for homework and exam situations, it
is naturally limited to the calculation capacities of Excel. There are
significantly lower than the ones of \textsc{R} and are thus not entirely
sufficient for exercises in higher statistics. Moreover, the approach requires
all students to have access to Excel, and teachers to apply a lot of security
measures to protect the spreadsheets against malicious attempts.

\subsection{General CAA Systems for Mathematics}
To overcome the limitations of tool support solely in exercises generation, CAA
systems for mathematics can be used. A well-known CAA system for mathematics is
\textsc{ActiveMath}. It allows to generate variable content within exercises
based on randomization
\cite{MelisEtAl2001ActiveMath,Goguadze2011ActiveMathInteractiveExercisesDomainReasoners}.
Randomization is realized by drawing a value from a set of admissible values for
each variable, where admissible values may be numbers (both from sets and
intervals), expressions, functions and alike. Further processing of the input is
possible via the Computer Algebra System (CAS) used within \textsc{ActiveMath}, but not by invoking an
external system such as \textsc{R}. The same is true for the systems
\textsc{Alice} \cite{Strickland2002Alice} and \textsc{MapleTA}
\cite{BlythLabovic2009MapleTA,Maple2015MapleTA}, which are based on
\textsc{Maple} as a CAS and thus naturally limited to the features offered by
that system. Consequently, there is no CAA system which already supports the
requirements sketched in the introduction to full extent.

\textsc{Wiris}\footnote{\url{http://www.wiris.com}} offers components to be
embedded into learning environments instead of being a CAA system on its own.
This could be a promising approach to add the required features to an existing
system. The combination of their quiz engine and CAS makes exercises
programmable to allow for randomization. In particular, loops and conditional
statements can be used to generate random content repeatedly until certain
desired properties are met. While this allows the exercise designer to add
virtually any functions not offered by the CAS directly, one looses the
performance and quality benefits of specialized computer algebra systems and
specialized software like \textsc{R}. Similar is true for tools executing
assessment items described in some standard format like
\textsc{QTIWorks}\footnote{\url{https://webapps.ph.ed.ac.uk/qtiworks/}} for
items defined in the QTI 2.1 format. In these cases, the combination of QTI 2.1
and the particular tool does not allow to use external software like \textsc{R}.

A general framework for math assessments supported by CAS is offered by the
\textsc{CABLE} framework \cite{NaismithSangwin2004ComputerAlgebraAssessment}.
While it allows to use virtually any CAS or external system to generate variable
values for exercise variants and to evaluate the student's input, its evaluation
is limited to algebraic expressions. Consequently, it offers similar support as
the tools for generating paper based exercises, but only little support in
automated evaluation and feedback. In particular, correctness of a solution is
determined by testing whether the difference between an input and the model
solution is zero. Hence exercises that cannot be assessed this way cannot be
designed in the \textsc{CABLE} framework.

\section{The e-assessment system JACK}
\label{sec:central_message}

As a general tool for e-assessment and automated tutoring, exercises in JACK are
not limited to the domain of mathematics. Instead, exercises may contain
multiple choice, fill-in, and drop-down elements for receiving input from the
students. Thus a minimal exercise consists of some sort of task description, at least
one element that receives input from students, and at least one feedback message.
However, JACK offers some options for a more sophisticated exercise design and
more detailed feedback messages, in particular for exercises with mathematical
content, including the support of \LaTeX, input with a formula editor and an
evaluation of solutions using computer algebra systems. We will discuss the options
in the following subsections. 

\subsubsection{Parameterization of exercises}
As a basic feature of JACK, tasks may be generic by using variables as
placeholders. These placeholders are filled in dynamically, so the exercise
presents different content to the student every time it is attempted. There is no
limitation in where to use these variables, so the task description may vary,
the number or content of multiple choice or drop-down options may vary, the
expected correct results may vary, and so on. While this is of no immediate use
for a single visit on a single exercise, it is very helpful when students work
with the tutoring system for a longer time. In this case, they can work on the
same exercise more than once, receiving different values within that exercise.
Thus exercises remain challenging for a longer period of time. Moreover, it
possibly helps students to understand the abstract concepts and encourages them
to talk to each other about solution strategies instead of plain solutions.
Parameterization even offers another benefit during summative assessments, as 
it helps to avoid plagiarism between students.
Using \textsc{R} to parameterize exercises enables us to use the wide range of 
statistical functions the system offers. In particular, data sets on the basis 
of arbitrary random variables can be generated, which is a very useful feature.
Moreover, JACK allows the use of plots created with \textsc{R}. A such plot can 
even be based on a randomly generated data set, which introduces another type 
of variable elements in an exercise: image variables.


\subsubsection{Feedback Options}
As a very important consequence of splitting an exercise into steps, the student
may receive detailed individual feedback for each step. A feedback in general
consists of a score and a feedback message. According to the typology of
Tunstall and Gsipps \cite{TunstallGsipps1996FeedbackTypology}, JACK thus provides both evaluative
and descriptive feedback: The score is an integer number in the range of $0$ to
$100$ based on a grading scheme provided by the author. Hence it is an
evaluative feedback that provides a judgment and tells the student whether he
was right or wrong. The feedback message may contain arbitrary content,
including dynamically created graphics. In particular, it can refer to the
student's input, reuse values from previous steps, and involve any kind of
calculation. Hence it is a descriptive feedback that refers to the student's
achievements and may provide guidance on how to improve a wrong solution.

We consider the latter kind of feedback as one of the central features of a
tutoring system. It is intended to help the students to comprehend where they made
mistakes or where they were correct. To be able to do so, JACK has to understand
the semantics of a solution. For this reason, CAS are
used to evaluate solutions. On one hand, the CAS can verify the correctness
of a solution, even if there are infinitely many correct ones. On the other hand, it is able
to locate errors precisely and to compare a student's solution with a standard
solution. This enables authors to provide granular feedback, evaluative feedback 
as well as descriptive feedback. Section \ref{sec:realization} explains 
in a more detailed way how external systems can be used to provide feedback.

\begin{figure}
	\centering
	\includegraphics[width=\columnwidth]{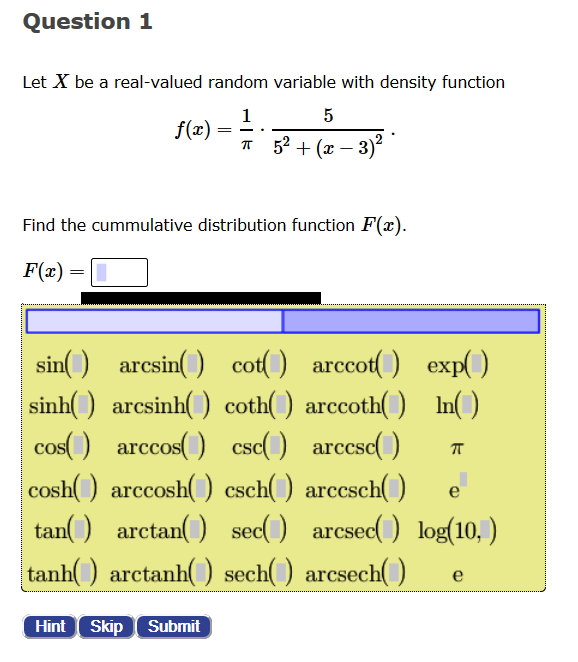}
\caption{\footnotesize Screen capture of the example exercise.}
\label{fig:feedbackexample}
\end{figure} 

Let us consider an example exercise to illustrate how the system is able to give feedback. 
In the first step of the exercise the student has to compute the cumulative distribution function $F(x)$ 
for a random variable given by its density function $$f(x) = \frac{1}{\pi} \cdot \frac{k}{k^2+(x-m)^2},$$
where $k$ and $m$ are randomly generated integers. The right way to solve this exercise is to compute the integral 
$$F(x) = \int_{- \infty}^{x}{f(t) dt} = \frac{1}{\pi} \cdot \arctan \left(\frac{x-m}{k} \right) + \frac{1}{2}.$$
As we can see in figure \ref{fig:feedbackexample}, the student submits his solution with the 
help of a formula editor, which provides many trigonometric and hyperbolic functions, which 
allows him to enter the $\arctan$-function into the input text field. In case the submitted 
solution is correct, the system tells the student so and takes him to step 2 of the exercise 
where he has to compute a quantile of the distribution. In addition to that, the author of 
the exercise has created a series of feedback messages for incorrect solutions. In that case, 
the student can use the feedback to improve his submission and redo the step. The feedback messages 
are the following:

\begin{enumerate}
  \item The system checks, whether the submitted solution depends on the required variable $x$. 
  If this is not the case, a feedback message is given, telling the student to use $x$.
  \item The system checks, whether the arctangent-function was used and provides a feedback message otherwise.
  \item Most students probably compute the integral by substituting $s = \frac{t-m}{k}$ or similar. 
   It can then happen that they forget the factor $\frac{1}{k}$ when replacing $dt$ by $ds$. This would lead to 
   the solution $F(x) = \frac{1}{\pi k} \arctan \left(\frac{x-m}{k}\right) + \frac{1}{2}$. So in case this 
   solution is submitted the system will provide a feedback message telling the
   student to recheck his substitution.
  \item If 1., 2. and 3. are not fulfilled the system checks whether the correct integration constant was used
   and provides a new step in case it is unequal to $\frac{1}{2}$. The system recapitulates how the integration 
   constant is determined and lets the student recompute it.
  \item In case the arctangent-function was used and the integration constant is correct, the system checks if 
  the argument of the arctangent-function is equal to $\frac{x-m}{k}$. A feedback message is given otherwise.
  \item When the checks mentioned in 1.-5. fail a default message is displayed, telling the student that his 
  solution is wrong.  
\end{enumerate}

A student who is not able to solve step 1 of the exercise can ask for a hint by clicking the button provided for this purpose. 
Authors can add multiple hints to a step which are shown separately, each time the student clicks the button.
For step 1 of this exercise the author has supplied three different hints. The first hint is very basic and recapitulates 
how to determine the cumulative distribution function from the density function. The second hint reminds the student how he 
can find the integration constant and the third hint tells him that $\arctan(x)$ is an antiderivative for $\frac{1}{1+x^2}$.
In none of these hints help the student to find the correct answer, he can use the skip button to move forward to step 2 
of the exercise. The system reveals the solution and tells the student how to find it.

\subsection{Further Examples}
\begin{figure*}
	\centering
	\includegraphics[width=\textwidth]{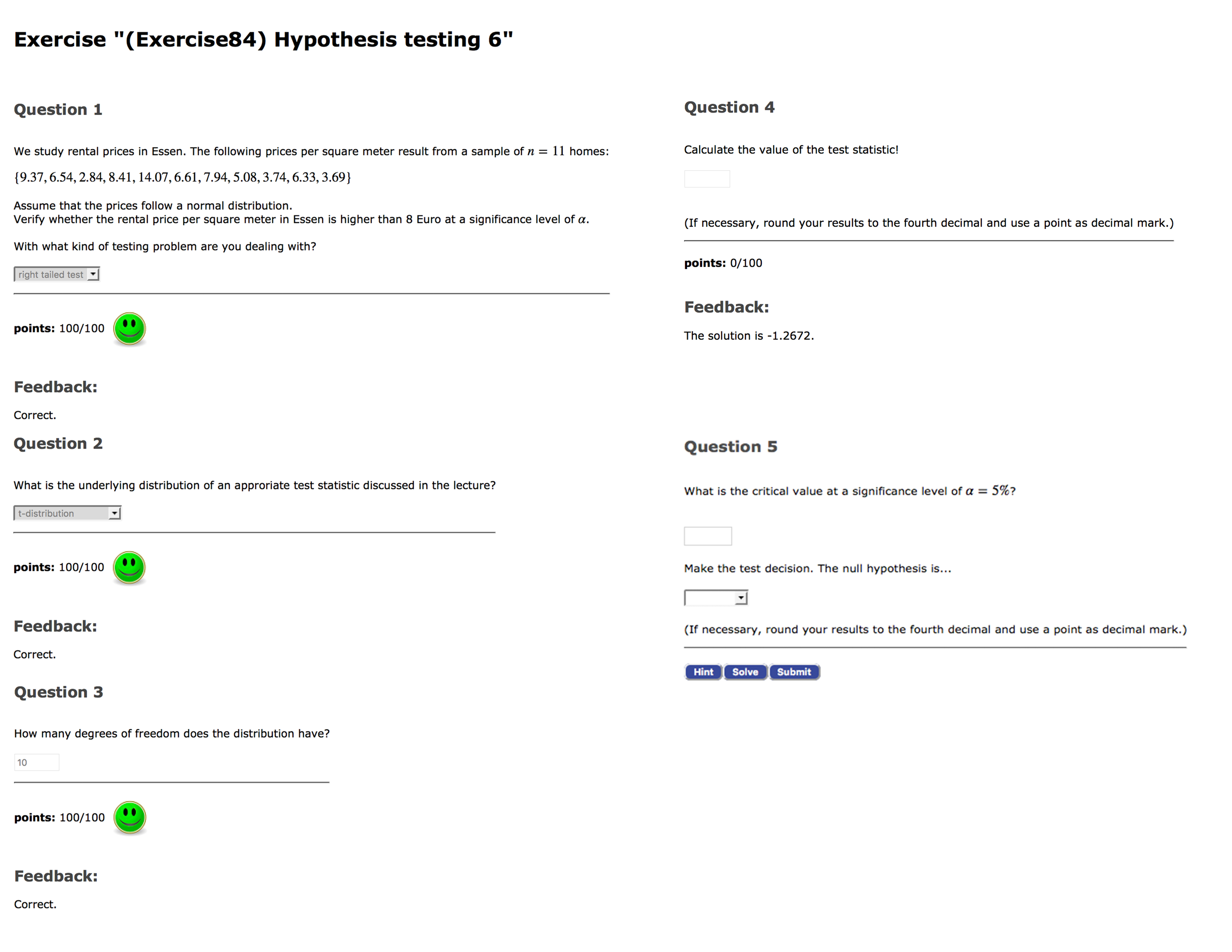}
\caption{\footnotesize Screen capture of a hypothesis testing exercise realized
in five stages.}
\label{pic11}
\end{figure*}

We will now present some examples from the lectures we used the system in to illustrate 
the steps that were necessary to convert them from paper-based exercises into digital ones.

The example in Figure \ref{pic11} introduces a more complex exercise type,
dealing with hypothesis testing. In a paper based test, exercises like this one
would require a longer answer including several arguments and results from
calculations. Hence it is not feasible to ask just for a single final result in
an electronic version of this exercise. Instead, we decompose the exercise into
five stages in JACK. On a technical level, each stage defines a stand-alone
exercise, containing a task, feedback messages, hints, etc. Of course, these
tasks are strongly related to each other by sharing the same context. The
decomposition into several stages enables us to provide detailed feedback
messages for each single task and to react differently to several possible flaws
in each of these stages.

In the first stage we draw a sample of raw data and ask if the question calls
for a left, right or two tailed alternative hypothesis in a drop-down menu.
Here, we use the possibility to draw more than one number from \textsc{R} at
once for the sample.

In case that the exercise is used for summative assessment it is advisable to
give partial credit for follow-up mistakes if only stage 1 is wrong.
Depending on the answer in this stage ("`right tailed"' is the correct
solution"') the exercise is path dependent. This means even if the first stage
is answered incorrectly the user can proceed on this subpath performing e.g.~a
left tailed test.

We then ask for the distribution of the test statistic, assuming the data to be
i.i.d.~normal, with some common distributions offered in the drop-down menu
(question 2) and for the degrees of freedom of the null distribution (question
3). Stage 3 is only visible if the student selects the Student t distribution in
stage 2. If the normal distribution is chosen the user is directed to stage 4 as
it now makes no sense to ask for the degree of freedom.

The choice of stages is not restricted to using drop-down menus. Hence we can
also proceed differently after stage 3, depending on the value typed into the
input field by the student. Notably, experience has shown that students do not
always submit numeric solutions. They may, for example, enter a ``big O''
instead of a zero. Hence, we also have to define a fallback stage in case the
submission is not numeric, because otherwise we cannot make any computations
with it in the next stage.

In stage 4 and 5 we finally ask for results from calculations, using two
strategies to allow minor deviations in the actual numbers: First, we ask for
the result in a precision of four decimals and thus allow to omit minor rounding
differences. Second, we configure the exercise to allow a corridor of input
around the precise correct solution. So if -1.2672 is correct as in stage 4,
input in the range from -1.2662 to -1.2682 is accepted as correct. The size of
the range can be adjusted individually for each stage.

In the sample session shown in Figure \ref{pic11}, the student skipped stage 4
and thus JACK shows the correct result instead of giving credits or other
feedback. Fallback stages as mentioned above are necessary for such cases as
well, as we cannot proceed with user input if a stage is skipped.

In summary, this example demonstrates how an exercise that offers a great amount
of freedom in its paper based version can be transformed into a digital version
offering almost the same amount of freedom for the students. In particular,
feedback can be given to flaws in different stages of the exercise and students
can continue also with wrong answers or even if they are not able to answer a
particular stage at all.

\begin{figure*}
	\centering
	\includegraphics[width=0.6\textwidth]{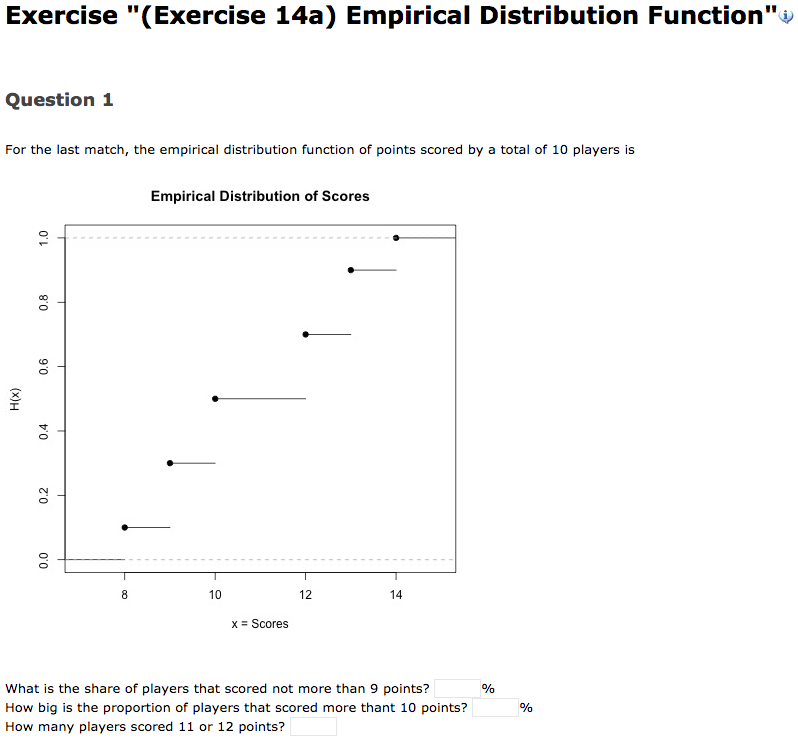}
\caption{\footnotesize Screen capture of an exercise using automated plot generation.}
\label{pic14a}
\end{figure*}

Another example using the graphics capabilities of \textsc{R} is shown in Figure
\ref{pic14a}. The plot used in the task description is created dynamically by
\textsc{R} based on random data following some pre-defined distribution. The
same random data is also used to compute the correct solution. There is not much
degree of freedom in the exercise, as it just asks for some simple numbers.
Nevertheless the exercise gets more interesting due to digitalization, as we are
able to produce virtually hundreds or thousands of different plots. This
includes drawing random data several times using the same distribution and
parameters as well as drawing random data using different parameters.

\section{Technical Realization}
\label{sec:realization}
The general framework for generic exercises in JACK as presented already in
previous publications 
needed no specific extensions to be used with sophisticated exercises for higher statistics.
However, the general architecture of JACK (as presented in \cite{Striewe2016ArchitectureModularGrading}) 
was not yet completely prepared for the flexible integration and very frequent use of external systems like
\textsc{R}. In particular, it is not feasible to simply start a new \textsc{R}
process on the server every time exercise content needs to be generated or
student input needs to be evaluated. This would first imply a large overhead on
system load with respect to starting and stopping processes and second would
make it difficult to run more complex commands for feedback generation such as
creating plot images based both on exercise parameters and student input.

To avoid these difficulties, the architecture of JACK has been extended as
follows: An instance of \textsc{Rserve}\footnote{\url{https://rforge.net/Rserve/}} is
deployed alongside JACK on the server which accepts multiple connections at once
and is able to process several commands issued over one of the connections one
after another. Each connection is associated with a dedicated workspace both in
memory and on the disk which keeps a persistent state as long as the connection
is not closed. When a student starts to work on an exercise, a new connection to
\textsc{Rserve} is opened specifically for this student and this exercise and
kept open until the student leaves the exercise. Hence there is less overhead
for starting and stopping \textsc{R} processes. Furthermore, results from
creating exercise parameter values or processing student input can be stored in
the workspace and reused later on in the same exercise when processing further
input or creating more parameter values. Once the student leaves the exercise,
the connection is closed and all workspace content is deleted. Consequently, the
student can start with completely new parameter values when trying the same
exercise again.


\section{Usage Scenarios and Evaluation}
\label{sec:in_use}

\begin{table*}
	\centering
	\begin{tabular}{c|c|c|c|c|c}
		\textbf{Course} & \multicolumn{3}{|c|}{\#Exercises} & \#Students & \#Submissions \\ \hline
		& practice & tests & exams & & \\ \hline
		Descriptive statistics & 92 & 22 & & 398 & 51803  \\ \hline
		Inducive statistics & 133 & 40 & 4 & 153 & 25949 \\ \hline
	\end{tabular}
	\caption{\footnotesize Usage figures for the exercises created for the use in the lectures of descriptive and inductive statistics}
	\label{table:summary}
\end{table*}

The University of Duisburg-Essen offers two large lectures for statistics that are supported with JACK. These are 
the courses for descriptive and inductive statistics held by the faculty of economics. Both are attended by up to 
700 students per semester. The concept of using CAA is the same in both courses. We use formative and summative 
assessments during the semester and even offered electronic exams as a voluntary alternative to the normal exam.
 
The formative assessments are intended to replace the classical paper-based exercises that students often have to 
do in traditional university courses with mathematical content.
Summative assessments during the semester are offered as small tests every other week. Students can gain bonus points for the final exam in each of the tests. 
These summative assessments shall motivate them to start learning earlier in the semester. The tests are taken from at home, 
which enables the participants to work on them collaboratively and to use all kinds of resources to help them. Therefore, the 
bonus points are only added to the results of those students, who have passed the exam.
Two electronic exams are offered at the end of the semester and complement two paper-based exams. Students are free to choose 
which type of exam they prefer. Experiences show that both types of exams are equally accepted by the students. 

The different ways of using exercises for different purposes has a direct effect on the exercise design. Exercises created for the formative assessment 
type need to be provided with detailed hints and feedback messages in order to guarantee a good learning outcome.
Stages that are not solved correctly either have to be repeated or the task can be skipped manually.
 
Exercises used in summative assessments 
need to fulfill different requirements than those used in formative assessments. Hints and feedback messages are not shown 
to the students during the test, so it is not necessary to create them. However, repetitions of stages should not be allowed 
in test exercises, since this behaviour could reveal information about the correctness of the submission. Furthermore, we 
have to make sure that the exercises are able to deal with consequential errors when students have to use their input 
from previous stages for calculations in the following ones. This feature needs to be used very carefully, because 
experiences show that submissions may not be of the expected type which can make it impossible to use them in further stages. 
For example, doing calculations with user input is impossible if a student just entered ``don't know'' instead of a number.
To handle these cases, the exercise author can then define a so-called fallback stage that is used in this case.

In addition to the requirements for summative assessments, exercises created for the electronic exams also require a lot of testing and considerations on how to grade them. 
We have to spend a good amount of time on predicting possible erroneous submissions that are still worth some points.

The teachers of the two lectures and their assistants have to create the exercises themselves, as they are the only ones having the 
required domain knowledge. We offer small workshops where we teach them how to deal with the system. Experiences show 
that they can start creating exercises themselves very quickly. Consequently, a large exercise pool has been created 
for both lectures even though creating a well tested complex exercise can take up to 8 hours of time. 
Table \ref{table:summary} gives a brief summary of the created exercises for the different assessment types and 
their usage numbers. Especially shortly before the summative assessments we can observe a lot of traffic on the system 
which shows that students take the opportunity to practice. As most of the exercises for the formative assessments are 
parameterized, students tend to do them more than once. Furthermore the participation numbers in the electronic exams are 
equal to the paper-based exams. As students are free to choose which exam they would like to take this shows the acceptance 
of the sytem among them. The data we received shows that those students who worked the most during the semester appear to have 
the best performances in the exams. Therefore, we intend to use the learning behaviour as a predictor for the outcome in the exam.
However, a detailed analysis is subject to further research. Nevertheless, all together we can conclude that our approach is heading into the right direction.

\section{Conclusions and Future Work}
\label{sec:conclusions}


In this paper we introduced a flexible exercise type that allows the digitalisation of 
complex exercises used in academic teaching. We worked out and implemented the requirements 
that have to be fulfilled in order to be able to do this properly and without loss of quality.
To illustrate how the concept can work in practice we used the example of statistics.
In particular, we connected our system to the domain specific software \textsc{R} and 
created a large number of exercises. Most of these exercises are parameterized, which we consider 
another huge benefit of using CAA. The fact that the exercises we created are used within courses 
for statistics at our university shows that our concept is feasible.
However, the presented approach can be applied to any other domain using complex exercise 
designs. The architecture of our system allows us to easily connect it to other expert systems,
comparable to \textsc{R}. We have seen how such a system can be used to evaluate submissions.

In the domain of statistics we see further challenges that need to be overcome. The mentioned 
courses also teach students how to use \textsc{R} to perform complex computations.
A new exercise type that we are working on will be able to grade these small \textsc{R} programs 
automatically and to offer detailled feedback messages. We will integrate this new exercise type 
into our concept of supporting the courses with CAA.

With respect to the technical realization, there is only one further action
planned so far: The \textsc{RServe} instance should be moved to a separate
server, so that it can be used from different JACK frontend or backend instances
at the same time. While this would require a more sophisticated load balancing
concept on this dedicated server to avoid overloading it with requests, this
would make it much easier to make new \textsc{R} features available for all JACK
instances at the same time. Load balancing could happen by using
\textsc{Docker}\footnote{\url{https://www.docker.com/}} to spawn several instances of
\textsc{Rserve} if necessary.

\bibliography{references}
\bibliographystyle{plain}

\end{document}